%% file: ms.tex
\documentclass{aip-cp}

\usepackage[numbers]{natbib}
\usepackage{graphicx}
\usepackage{color}

\usepackage{wrapfig}

\usepackage{style_wei_aastex_no-bibstyle}

\usepackage{style_wei_symbols}
\usepackage{style_wei}
\usepackage{style_wei_journal_names}
  \renewcommand{\lrsp}{{Liv.~Rev.~Solar Phys.}}

\renewcommand*{\fig}[1]{Fig.~\ref{#1}}

\begin{document}

\title{Quasi-periodic Fast-mode Magnetosonic Wave Trains Within Coronal Waveguides 
Associated with Flares and CMEs}

\author[aff1,aff2]{Wei Liu\corref{cor1}}

\author[aff3,aff4]{Leon Ofman}
\author[aff6]{Brittany Broder}
\author[aff7]{Marian Karlick\'{y}}
\author[aff5]{Cooper Downs}

\affil[aff1]{Bay Area Environmental Research Institute, 625 2nd Street, Suite 209, Petaluma, CA 94952, USA}
\affil[aff2]{Lockheed Martin Solar and Astrophysics Laboratory, 
  3251 Hanover Street, Bldg.~252, Palo Alto, CA 94304, USA}	
\affil[aff3]{Catholic University of America and NASA Goddard Space Flight Center, Code 671, Greenbelt, MD 20771, USA}
\affil[aff4]{Visiting, Department of Geosciences, Tel Aviv University, Tel Aviv 69978, Israel}
\affil[aff6]{Department of Physics and Astronomy, Western Kentucky University, 	
  Bowling Green, KY 42101-1077, USA}
\affil[aff7]{Astronomical Institute of the Academy of Sciences, Ond\v{r}ejov, Czech Republic}
\affil[aff5]{Predictive Science Inc., 9990 Mesa Rim Road, Suite 170, San Diego, CA 92121, USA}

\corresp[cor1]{Corresponding author: weiliu@lmsal.com}

\maketitle

\vspace{-0.1in}

\begin{abstract}
Quasi-periodic, fast-mode, propagating wave trains (QFPs) 
are a new observational phenomenon recently discovered in the solar corona by
the {\it Solar Dynamics Observatory} with extreme ultraviolet (EUV) imaging observations. 
They originate from flares and propagate 
at speeds up to $\sim$$2000 \kmps$ within funnel-shaped waveguides 
in the wakes of coronal mass ejections (CMEs). 
QFPs can carry sufficient energy fluxes required for coronal heating during their occurrences.
They can provide new diagnostics for the solar corona and their associated flares.
We present recent observations of QFPs	
focusing on their spatio-temporal properties, temperature dependence,
and statistical correlation with flares and CMEs. 
Of particular interest is the 2010-Aug-01 C3.2 flare with correlated QFPs 
and drifting zebra and fiber radio bursts, which might be different manifestations of 
the same fast-mode wave trains. We also discuss the potential roles of QFPs in 
accelerating and/or modulating the solar wind.

\end{abstract}

\vspace{-0.15in}
\section{INTRODUCTION}
\label{sect_intro}
\vspace{-0.1in}


The dynamic, magnetized solar corona hosts a variety of plasma or magnetohydrodynamic (MHD) waves
that are believed to play important roles in many fundamental, yet enigmatic processes,
such as energy transport, 	
coronal heating, 	
and solar-wind acceleration.
%
Flare-associated, Quasi-periodic, Fast-mode Propagating 	
{wave trains} 	
\citep[{\bf QFPs}; see \fig{QFP-overview.eps}b;][]{LiuW.AIA-1st-EITwave.2010ApJ...723L..53L, LiuW.FastWave.2011ApJ...736L..13L, LiuW.cavity-oscil.2012ApJ...753...52L})
are a new, spectacular 	
coronal wave phenomenon discovered in extreme ultraviolet (EUV) by the Atmospheric Imaging Assembly (AIA)
onboard the {\it Solar Dynamics Observatory} (\sdo), thanks to its unprecedented high spatio--temporal resolution.
QFPs have been reproduced in MHD simulations and identified as propagating 
(as opposed to standing) fast-mode magnetosonic waves 
\citep[e.g.,][]{Ofman.Liu.fast-wave.2011ApJ...740L..33O, 
Pascoe.wing.QFPs.funnel.2D.MHD.2013A&A...560A..97P, 
PascoeDavid.coronal.hole.anti-wave-guide.QFPs.2014A&A...568A..20P,
YangLiping.QFPs.plasmoid.simul.2015ApJ...800..111Y}.

The {\bf significance} of QFPs lies in their potential novel diagnostics,
previously unavailable due to instrumental limitations.
First of all, they are intimately associated with solar flares 
during and even after coronal mass ejections (CMEs) and often originate from flare kernels.
They share some (but not all) periodicities with quasi-periodic pulsations 
(see \fig{QFP-overview.eps}c and \ref{QFP-overview.eps}d)
of their accompanying flares traditionally detected in non-imaging data from radio to hard X-rays
\citep[e.g.,][]{YoungCW.flare.radio.QPP.1961ApJ...133..243Y,	
Nakariakov.Melnikov.QPP.2009SSRv..149..119N}.
QFPs can thus provide critical insights to the 	
poorly understood mechanisms of energy release and associated plasma heating and particle acceleration processes
in flares, a fundamental question in solar physics. 	
Secondly, QFPs can serve as a new tool for coronal seismology
\citep[e.g.,][]{Uchida.coronal-seismology.1970PASJ...22..341U, 
Roberts.coronal-seismology.1984ApJ...279..857R}
to probe the physical properties of 
the solar corona, the medium in which they propagate. 	
For example, their funnel-shaped paths indicate the presence of waveguides, 
which provide a unique way to map the spatial distribution
of the fast-magnetosonic speed and thus the coronal magnetic field strength.
Thirdly, learning about QFPs can help us better understand wave propagation
in the solar corona and MHD turbulence in general.

QFPs are not uncommon. 	
A handful of them have been reported in the first five years of the \sdo\ mission
\citep[e.g.,][]{LiuW.AIA-1st-EITwave.2010ApJ...723L..53L, LiuW.FastWave.2011ApJ...736L..13L, LiuW.cavity-oscil.2012ApJ...753...52L,
ShenYD.LiuY.QFP.wave.2012ApJ...753...53S, ShenYD.LiuYu.QFP.171.193.2013SoPh..288..585S,
YuanD.QFP.distinct.trains.2013A&A...554A.144Y, NisticoG.2013Dec7.M1.2.QFP.2014AA...569A..12N,
ZhangYuZong.fast-slow-modes.2015A&A...581A..78Z}.
The general properties of QFPs were reviewed by \citet{LiuW.OfmanL.EUV.wave.review.2014SoPh..289.3233L}.
Yet, our knowledge of this new phenomenon is still rudimentary.
In this paper, we report recent progress in understanding QFPs,
highlight a few interesting findings, and present a preliminary survey of QFPs
and their flare/CME association.


\vspace{-0.16in}
\section{SPATIO\,--\,TEMPORAL PROPERTIES}	
\label{sect_Aug01}
\vspace{-0.12in}

\begin{figure}[thbp]      
     \centerline{
      \includegraphics[height=4.6cm]{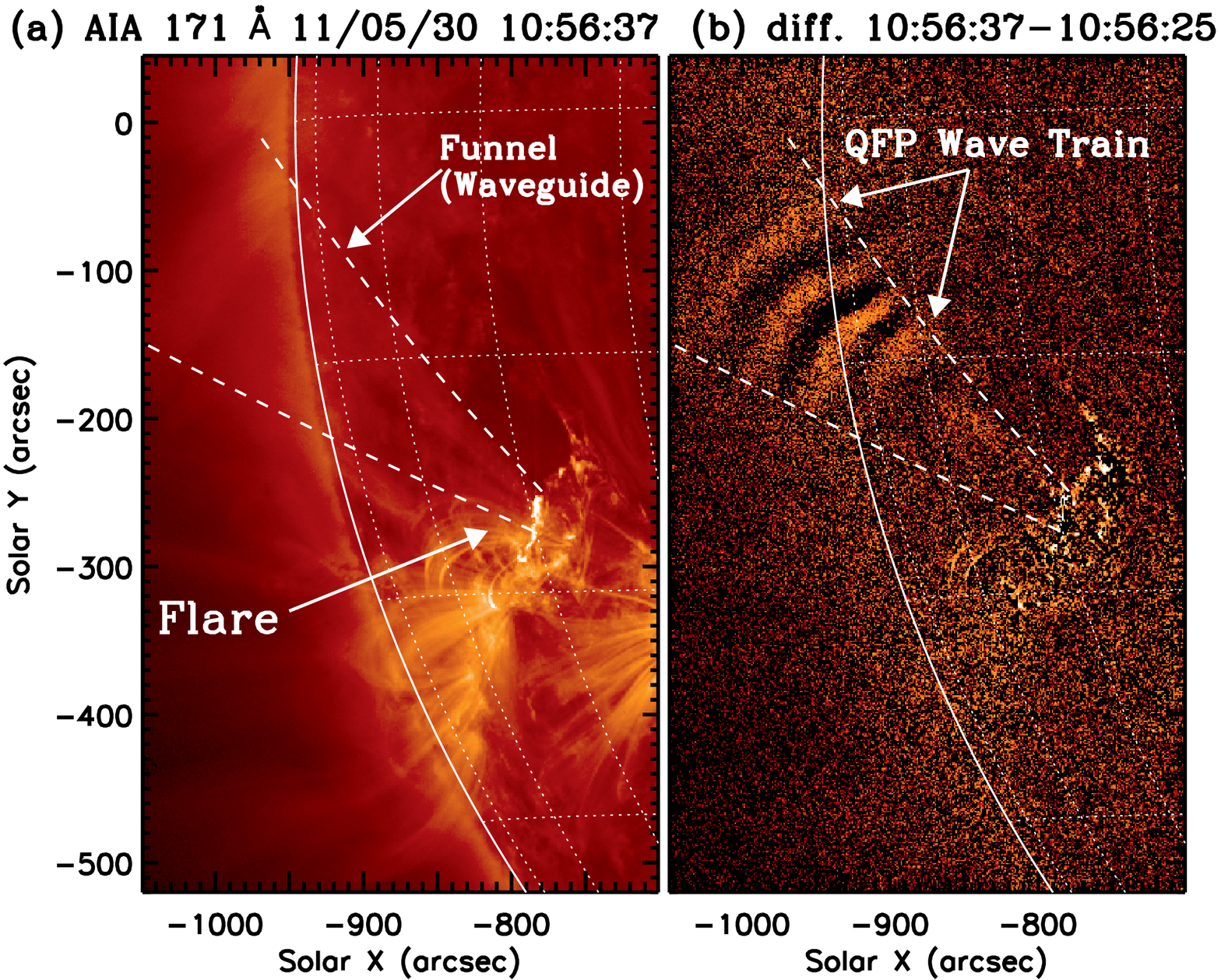}
      \includegraphics[height=4.6cm]{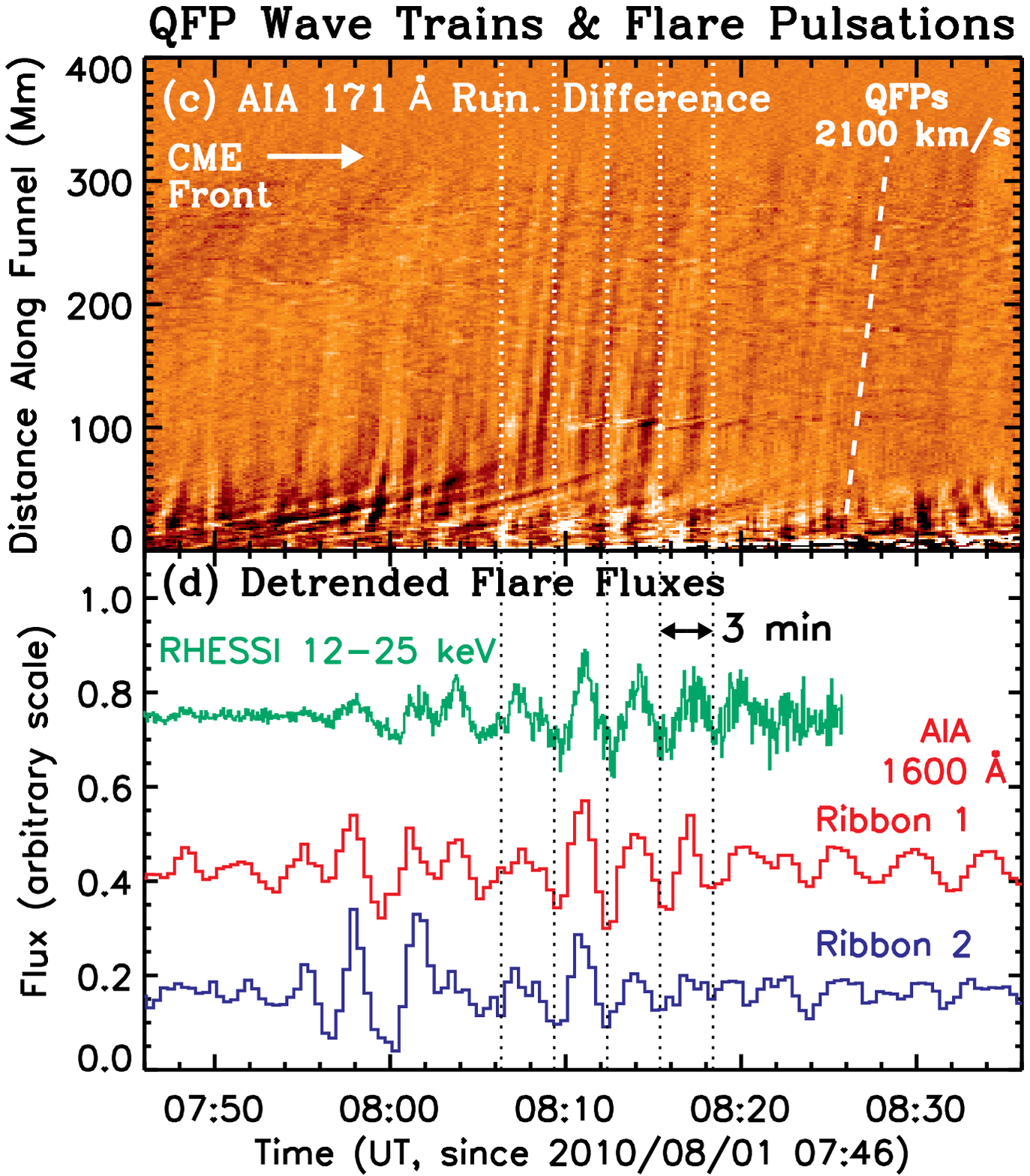}
	  \includegraphics[height=4.6cm]{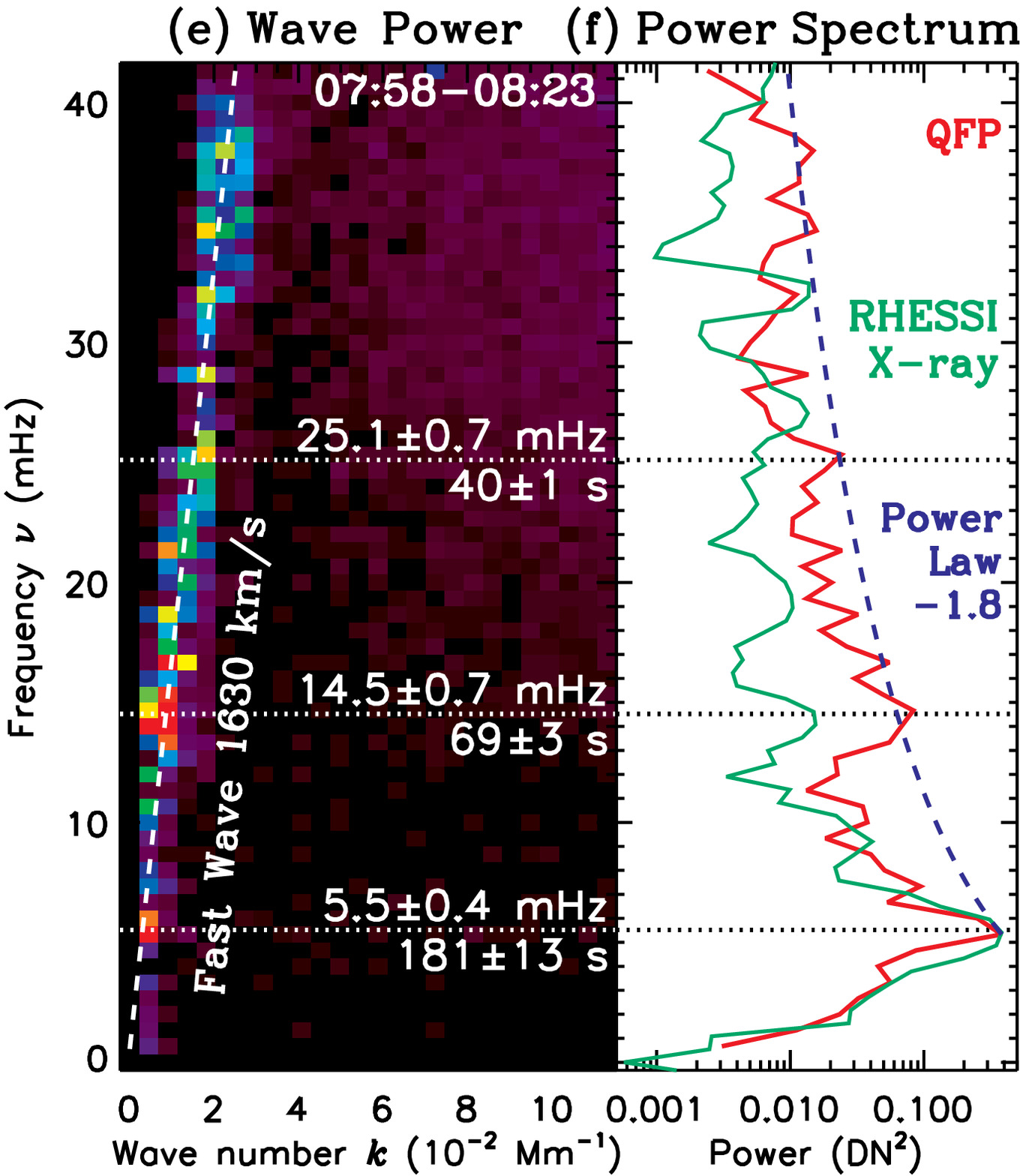}
      \includegraphics[height=4.6cm]{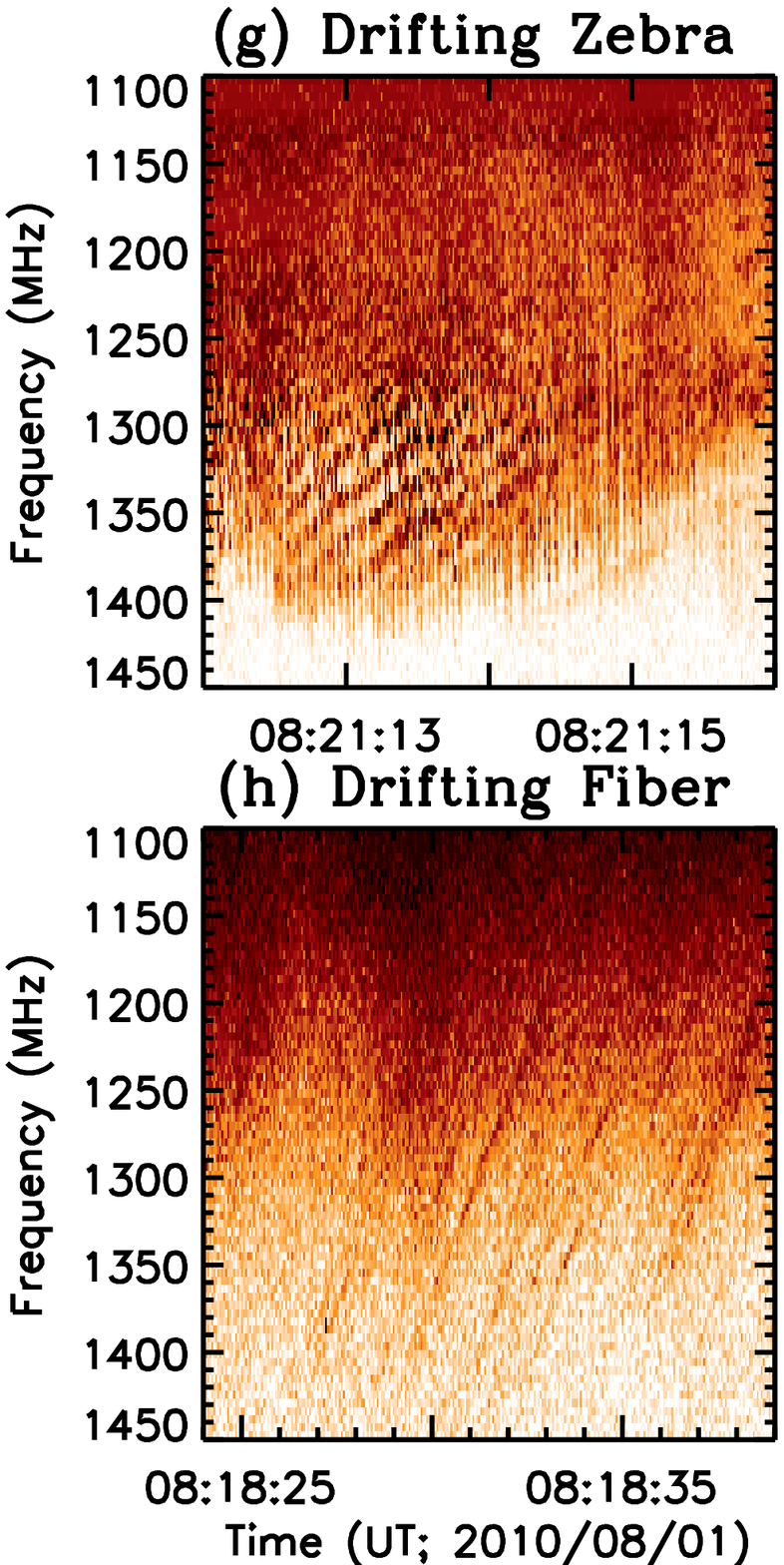}
     }
 \caption{
   {\bf Left}: Example of QFPs in a funnel (waveguide) rooted at the 2011-May-30 C2.8 flare
  shown in (a) AIA 171~\AA\ direct and (b) difference images, which were analyzed by
 \citet{ShenYD.LiuY.QFP.wave.2012ApJ...753...53S} and \citet{YuanD.QFP.distinct.trains.2013A&A...554A.144Y}.
   The rest of this figure are for the 2010-Aug-01 C3.2 flare event.
   {\bf Middle} (modified from \citet{LiuW.FastWave.2011ApJ...736L..13L}):
 (c) QFPs shown in a 171~\AA\ space--time plot and (d) their correlated
 X-ray and UV pulsations 	
 at a dominant 3-min period.
  (e) Fourier power or $k$--$\omega$ diagram of QFPs shown as
 a bright ridge and (f) wave-number averaged wave power as a function of frequency.
    {\bf Right}: (g) Zebra and (h) fiber radio bursts
 detected by the Ond\v{r}ejov Observatory radiospectrograph
 \citep[data from][]{KarlickyM.zebra.my-2010Aug01-QFP.2014AA...561A..34K}.		
 } \label{QFP-overview.eps}
 \end{figure}

{\bf Frequency Distributions and Power Spectra:} 
QFPs are generally observed in a wide period range from 25
to $\sim$400~s, with the lower end limited by the Nyquist frequency 	
given by AIA's 12-second cadence. In Fourier power spectra or $k$--$\omega$ diagrams
(e.g., \fig{QFP-overview.eps}e),
QFPs appear as bright, nearly straight ridges,	
which describe their dispersion relations. 
Individual peaks of power on the ridge are often concentrated within a period range of 40\,--\,240~s. 

In the heavily studied 2010-Aug-01 C3.2 flare/QFP event 
\citep[e.g.,][]{Schrijver.Title.2010Aug01-long-range-couple.2011JGRA..116.4108S}, 
for example, \citet{LiuW.FastWave.2011ApJ...736L..13L}
found that such power peaks were distributed
approximately in a power law of frequency $\nu$ with an index of $-(1.8 \pm 0.2)$,
as shown in \fig{QFP-overview.eps}f.
This index, close to the Kolmogorov value of $-5/3$ for turbulence,	
is similar to that of the zebra radio bursts detected in the same event
\citep[][his Fig.~6]{KarlickyM.zebra.my-2010Aug01-QFP.2014AA...561A..34K}
and those found elsewhere in the corona		
\citep[][their Table~1]{TomczykMcIntosh.coronal-time-distance-seism.2009ApJ...697.1384T,
IrelandJ.corona.power.spec.2015ApJ...798....1I}.
This suggests that QFPs could be part of the 
ensemble of wave turbulence permeating and potentially heating the corona.


The detection of zebra and fiber radio bursts	
in the same 2010-Aug-01 event
\citep{KarlickyM.zebra.my-2010Aug01-QFP.2014AA...561A..34K},
as shown in \fig{QFP-overview.eps}g and \ref{QFP-overview.eps}h
bears further interesting implications.
Their drifts from high to low frequencies with time
were ascribed to coherent plasma emission modulated by 
upward propagating fast-mode magnetosonic waves
\citep{KarlickyM.fiber.burst.fast.sausage.wave.train.2013A&A...550A...1K,
KarlickyMarian.zebra.mudolated.by.fastMode.2013A&A...552A..90K}.
Assuming an atmospheric density model \citep[][p.~188]{AschwandenM2002SSRv..101....1A},
the frequency drifts indeed yield propagation speeds of $\sim$$1000 \kmps$, 
which are roughly consistent with those AIA-detected QFPs
and thus suggest a common origin of fast-mode wave trains.
%
%
\vspace{-0.1in}
\vspace{-0.1in}
%

\vspace{-0.16in}
\section{TEMPERATURE DEPENDENCE}	
\label{sect_Apr08}
\vspace{-0.12in}

QFPs are best (and often only) detected in AIA's 171~\AA\ channel
and occasionally in the 193 and 211~\AA\ channels. 	
Possible underlying reasons 	
are two-fold:
(1) The wave-hosting plasma is likely near the 171~\AA\ channel's peak response temperature of $\sim$$0.8 \MK$,
rather than those of the 193 and 211~\AA\ channels ($\sim$1.6 and $2.0 \MK$, respectively).
In addition, QFPs, unlike global EUV waves of often large amplitudes
\citep[e.g.,][]{DownsC.MHD.2010-06-13-AIA-wave.2012ApJ...750..134D}, 
cause smaller perturbations (to the plasma) and thus too small temperature departures
to appear in other AIA channels.
(2) The 171~\AA\ channel has a much higher photon response efficiency 
than any other AIA channel by at least one order of magnitude. 
Therefore it is particularly sensitive to small intensity variations, 
typically on the 1--5\% level for QFPs.

\begin{figure}[thbp]      
 \centerline{
 \includegraphics[height=2.in]{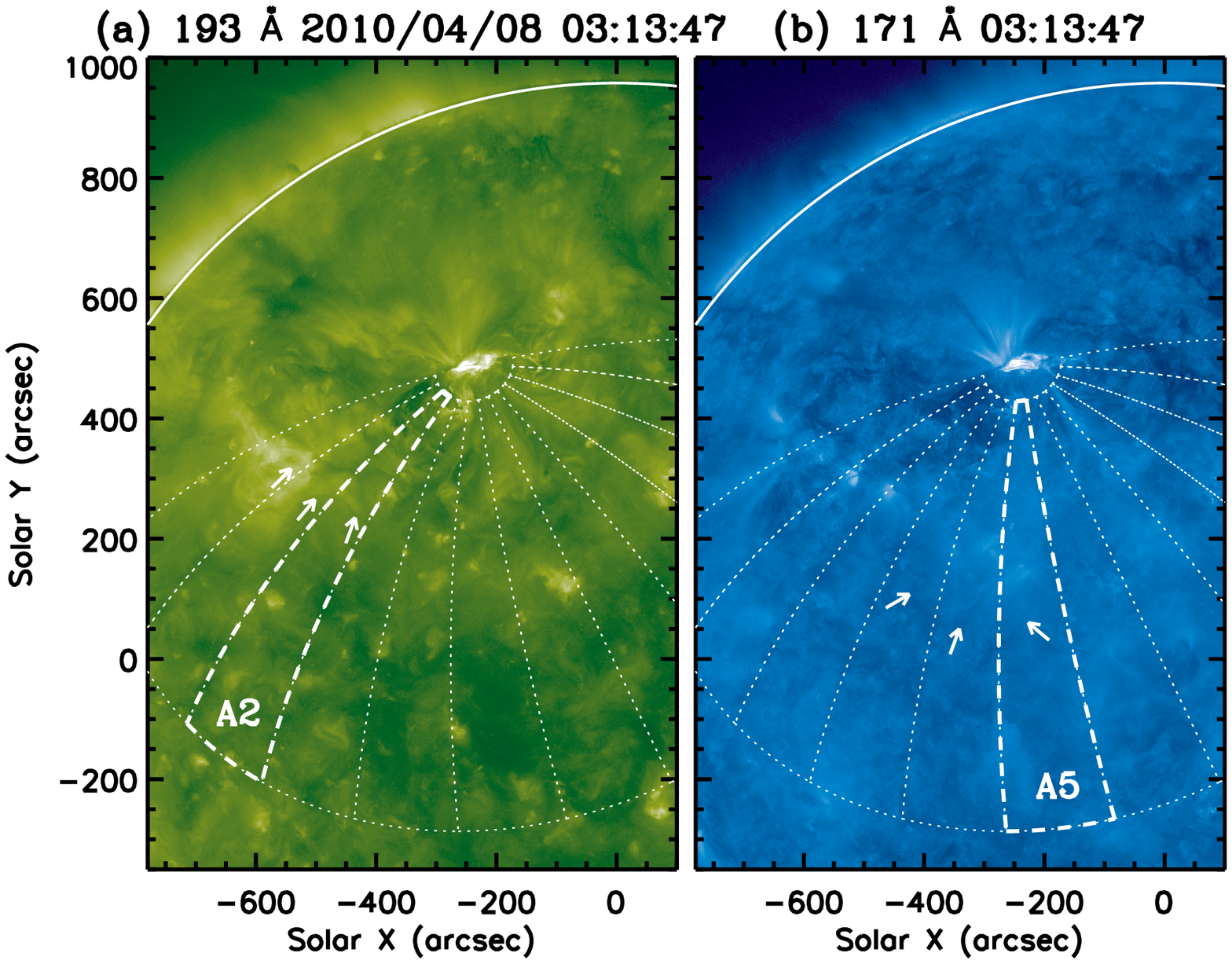}	
 \includegraphics[height=2.in]{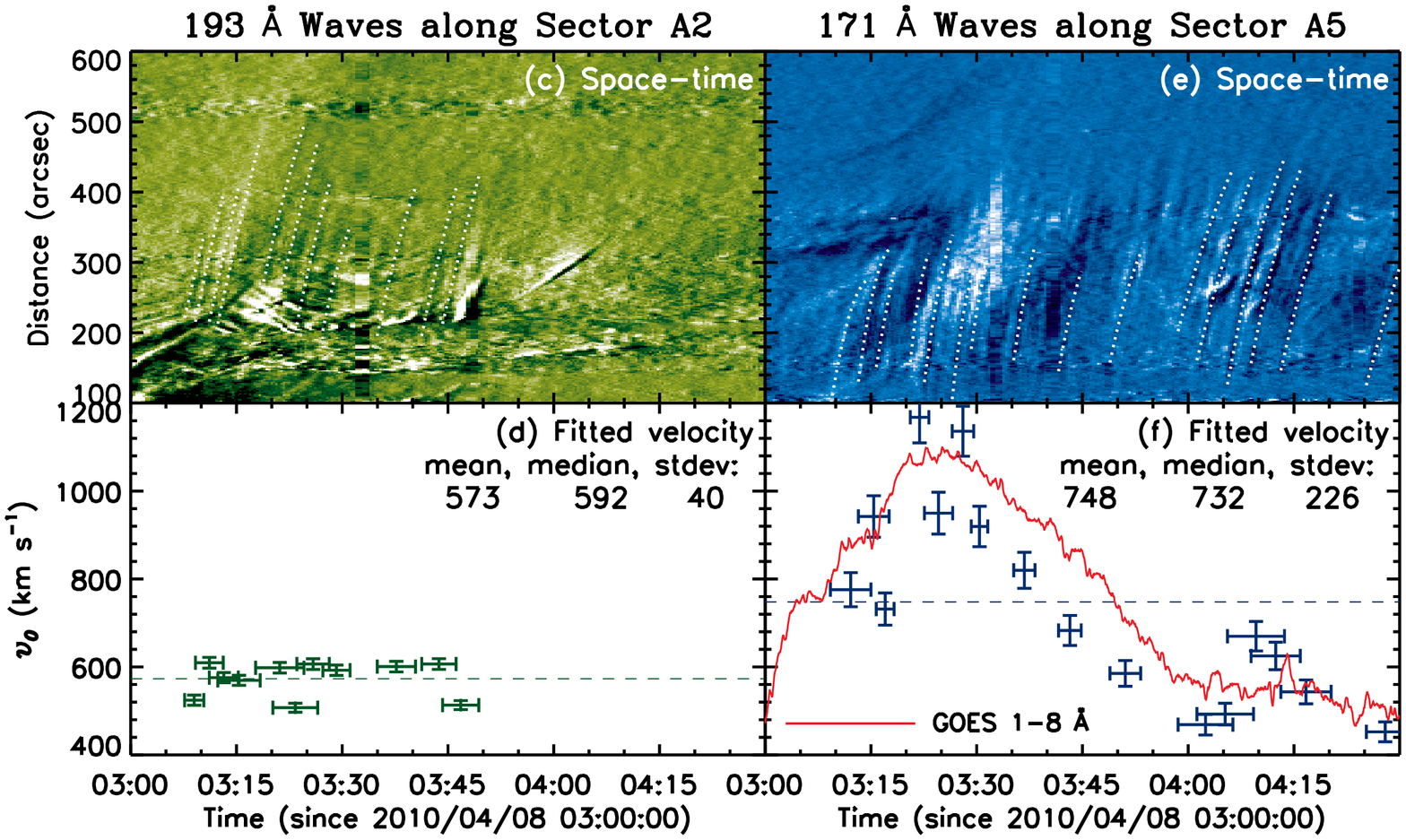}
 }
\vspace{-0.10in}
 \caption{Example of the temperature dependence of QFPs during the 2010-Apr-08 B3.7 flare,
whose associated CME eruption \citep{SuYN.flux.rope.insert.CME.2010Apr08.2011ApJ...734...53S}
produced the first global EUV wave \citep{LiuW.AIA-1st-EITwave.2010ApJ...723L..53L}
and non-linear Kelvin-Helmholtz instability waves
\citep{	
Ofman.Thompson.AIA.KH.instab.2011ApJ...734L..11O}
detected by \sdo/AIA shortly after its first light.
  (a) \& (b)~AIA 193 and 171~\AA\ images overlaid with spherical sectors centered at the source flare
 (modified from \citep{LiuW.AIA-1st-EITwave.2010ApJ...723L..53L}).	
 The arrows mark the locations of QFP wave fronts.
  (c)~Running-ratio space-time plots at 193~\AA\ obtained from Sector~A2 shown in (a).
 The white dotted lines are parabolic fits to identified QFP wave trains.
  (d)~The initial wave speeds from the fits in (c) as a function of time.
 The horizontal error bars show the durations of the fits. 
 The horizontal dashed line indicates the mean velocity.
  (e) \& (f) Same as (c) \& (d) but for 171~\AA\ waves within Sector~A5 shown in (b).
 The red curve is the \goes~1--8~\AA\ flux of the flare 
 showing an interesting temporal correlation with the initial wave speeds.
  } \label{QFP-2trains.eps}
 \end{figure}

QFPs detected at 171~\AA\ and 193/211~\AA, occasionally in the same event,
often have considerable differences, e.g., in speed and spatial domain
\citep[][]{LiuW.cavity-oscil.2012ApJ...753...52L, ShenYD.LiuYu.QFP.171.193.2013SoPh..288..585S}.
171~\AA\ waves usually appear closer to the source flare and propagate at
higher speeds than 193/211~\AA\ waves.
This is consistent with the rapid decrease of the fast-magnetosonic speed away from 
the active region core. Such waves can also appear in different directions,
consistent with the inhomogeneous temperature distribution of 
plasma in different spatial regions, which was recently 
verified in MHD simulations of QFPs (Downs et al. 2015, in prep.).
An example of such distinct behaviors is shown in \fig{QFP-2trains.eps} for 
the 2010-Apr-08 B3.7 flare/QFP event.	
The strongest 171~\AA\ wave trains (blue) are separated by $\sim$$45 \degree$
in propagation direction from their 193~\AA\ counterparts (green) and have
a mean initial speed of $750 \kmps$ vs.~$570 \kmps$.
%

\vspace{-0.1in}
\section{STATISTICAL SURVEY OF QFPS}	
\label{sect_survey}
\vspace{-0.13in}

\begin{wrapfigure}{r}{0.6\textwidth} \vspace{-7pt}
 \centerline{
  \includegraphics[width=0.6\textwidth]{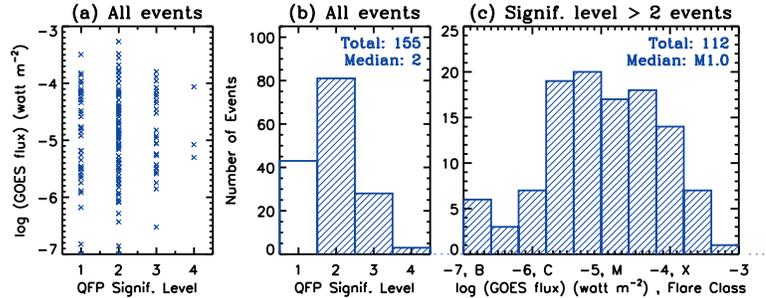}
 }
 \vspace{-10pt}
 \caption{Distribution of identified QFP events during 2010/06\,--\,2014/12.
    (a) \goes~1--8~\AA\ channel flux vs.~QFP significance level~$S$, showing no clear correlation. 
    (b) \& (c) Histograms of QFP events distributed in $S$ and flare classes.
  Only $S \geq 2$ events, considered as definitive detection of QFPs, are shown in (c)
  with a median flare class of M1.0.
   } \label{FPs-stat.eps}
 \vspace{-7pt} \end{wrapfigure}

Recently, we performed a preliminary survey of QFPs 	
and found that they were rather common.
We exhaustively scanned global EUV waves from June 2010 to December 2014 during the first 4.5 years of the \sdo\ mission,
which are currently cataloged at LMSAL (\citealt{NittaN.AIA.wave.stat.2013ApJ...776...58N}; 
{http://www.lmsal.com/nitta/movies/AIA\_Waves}).
Out of the 355 global EUV waves, we identified 155 preliminary QFP events.
We then assigned each event a significance level~$S$ in the range of 1--4,
depending on the wave amplitude (contrast), spatial size, and duration. 
Considering $S$$\geq$2 as definitive detection, we found 112 QFP events,
which translate to an association rate of $112/355 \approx 1/3$ with global EUV waves
that are all associated with CMEs and flares.
Figure~\ref{FPs-stat.eps} shows the distributions of 
their significance levels and flare classes,
with median values of level~2 and class~M1.0, 
but with no clear correlation between them.	

%
%
From another preliminary survey of flares from selected active regions, 
we found an interesting trend of preferential association of QFPs with	
{\it eruptive} flares, i.e., those accompanied by CMEs, rather than {\it confined} flares without CMEs.
QFPs can also cluster as recurrent, homologous events from certain ARs with favorable conditions, 
while rarely happen in some other ARs.
For example, 	
the recent record-setting, flare-rich AR~12192
produced almost no CMEs \citep{SunXudong.AR12192.flareBig.CMEpoor.2015ApJ...804L..28S, 
ChenHuadong.AR12192.confined.flares.2015ApJ...808L..24C}
and only one of its 26 flares inspected produced observable QFPs.
In contrast, AR~12205 produced 50\% less flares in total but
had five times more detectable QFP events (all with CMEs).	 
This suggests that a CME may create favorable conditions for QFP production or detection,
e.g., by forming a waveguide within funnel-shaped CME wakes 
as those detected in white-light eclipse images \citep{HabbalS.cone.wake.2011ApJ...734..120H}.	

\vspace{-0.15in}
\section{DISCUSSION}
\label{sect_discuss}
\vspace{-0.1in}

It has been proposed long ago that \Alfven\ waves as well as fast-mode magnetosonic waves 
can provide the momentum input and energy flux that accelerates/heats the solar wind 
\citep[e.g.,][]{AlazrakiG.Alfven.acc.solar.wind.1971A&A....13..380A,
FlaT.fast.mode.CH.solar.wind.1984ApJ...280..382F}.
These waves are particularly important for solar wind acceleration in open magnetic field regions such as
coronal holes \citep[e.g.,][]{ParkerE.phase.mixing.CH.SW.1991ApJ...376..355P}	
and have been the subject of extensive numerical modeling 
\citep[see the review by][]{Ofman.solarwind-review.2010LRSP....7....4O}.
The QFP wave-trains discovered by \sdo/AIA and identified as fast magnetosonic waves with
sufficient energy flux to heat active regions \citep{Ofman.Liu.fast-wave.2011ApJ...740L..33O}
provided the first evidence for this mechanism in the lower corona
that can power the solar wind. These waves are closely related to 
\Alfven\ waves in low-beta plasma, and can carry large energy
fluxes due to their high wave speeds. 
Nevertheless, the exact occurrence rate of these waves and the contribution of 
undetected small scale events to the integrated continuous wave energy flux 
in the corona are currently unknown and will be subjects of
future observational and numerical studies.

Another outstanding question regarding QFPs is the origin of their periodicities. 
The same periods found in simultaneous QFPs and 	
flare pulsations suggest a yet to be determined common origin, which could be 
(1) pulsed energy release intrinsic to magnetic reconnection,	
such as repetitive ejections of plasmoids \citep[e.g.,][]{LiuW.cusp.flare.20120719M7.2013ApJ...767..168L,
YangLiping.QFPs.plasmoid.simul.2015ApJ...800..111Y}
or the flow-induced Kelvin-Helmholtz instability in the current sheet \citep{OfmanSui.HXR-oscil.2006ApJ...644L.149O},
or (2) MHD oscillations due to resonance or dispersion, such as three-minute chromospheric sunspot oscillations
at the same period of the dominant signal in some QFPs 
\citep{LiuW.FastWave.2011ApJ...736L..13L},
which may be related to the upward leakage of slow-mode magnetosonic waves		
\citep[e.g.,][]{	
DeMoortelI.3-5min.slow-mode.leak.2corona.2002A&A...387L..13D}.
 Such mode conversion or coupling is expected to occur in the chromosphere where the plasma-$\beta$
is close to unity \cite[e.g.,][]{Bogdan.coronal-wave-simul.2003ApJ...599..626B},
and could be potentially related to those fast-propagating sunspot waves recently detected
at the photospheric level \citep{ZhaoJunwei.sunspot.waves.2015ApJ...809L..15Z}.
Such possibilities remain to be verified in future investigations,
e.g., with joint observations by \sdo/AIA, \iris, \hinode, and DKIST among other
space missions or ground-based facilities.

\vspace{-0.13in}
\section{ACKNOWLEDGMENTS}
\vspace{-0.10in}

This work was supported by NASA LWS grant~NNX14AJ49G to PSI
and NASA contract~NNG09FA40C (\iris) to LMSAL.
W.~L.~was supported in part by NASA grants~NNX13AF79G and NNX14AG03G,
L.~O.~by NASA grant NNX12AB34G, NSF grant AGS 1059838 and NASA cooperative agreement NNG11PL10A to CUA,
B.~B.~by an \iris\ summer internship at LMSAL and Stanford University,
and M.~K.~by Grant~P209/12/0103~(GA~CR).

\bibliographystyle{aipnum-cp}
\vspace{-0.09in}
\input{ms_edit.bbl}
\end{document}

%% file: ms_edit.bbl
%

%% file: ms.bbl
\begin{thebibliography}{39}%
\makeatletter
\vspace{-0.13in}
\providecommand \@ifxundefined [1]{%
 \@ifx{#1\undefined}
}%
\providecommand \@ifnum [1]{%
 \ifnum #1\expandafter \@firstoftwo
 \else \expandafter \@secondoftwo
 \fi
}%
\providecommand \@ifx [1]{%
 \ifx #1\expandafter \@firstoftwo
 \else \expandafter \@secondoftwo
 \fi
}%
\providecommand \natexlab [1]{#1}%
\providecommand \enquote  [1]{``#1''}%
\providecommand \bibnamefont  [1]{#1}%
\providecommand \bibfnamefont [1]{#1}%
\providecommand \citenamefont [1]{#1}%
\providecommand \href@noop [0]{\@secondoftwo}%
\providecommand \href [0]{\begingroup \@sanitize@url \@href}%
\providecommand \@href[1]{\@@startlink{#1}\@@href}%
\providecommand \@@href[1]{\endgroup#1\@@endlink}%
\providecommand \@sanitize@url [0]{\catcode `\$12\catcode `\&12\catcode
  `\#12\catcode `\^12\catcode `\_12\catcode `\%12\relax}%
\providecommand \@@startlink[1]{}%
\providecommand \@@endlink[0]{}%
\providecommand \url  [0]{\begingroup\@sanitize@url \@url }%
\providecommand \@url [1]{\endgroup\@href {#1}{\urlprefix }}%
\providecommand \urlprefix  [0]{URL }%
\providecommand \Eprint [0]{\href }%
\providecommand \doibase [0]{http://dx.doi.org/}%
\providecommand \selectlanguage [0]{\@gobble}%
\providecommand \bibinfo  [0]{\@secondoftwo}%
\providecommand \bibfield  [0]{\@secondoftwo}%
\providecommand \translation [1]{[#1]}%
\providecommand \BibitemOpen [0]{}%
\providecommand \bibitemStop [0]{}%
\providecommand \bibitemNoStop [0]{.\EOS\space}%
\providecommand \EOS [0]{\spacefactor3000\relax}%
\providecommand \BibitemShut  [1]{\csname bibitem#1\endcsname}%
\let\auto@bib@innerbib\@empty
\bibitem [{\citenamefont {{Liu}}\ \emph {et~al.}(2010)\citenamefont {{Liu}},
  \citenamefont {{Nitta}}, \citenamefont {{Schrijver}}, \citenamefont
  {{Title}},\ and\ \citenamefont
  {{Tarbell}}}]{LiuW.AIA-1st-EITwave.2010ApJ...723L..53L}%
  \BibitemOpen
  \bibfield  {author} {\bibinfo {author} {\bibfnamefont {W.}~\bibnamefont
  {{Liu}}}, \bibinfo {author} {\bibfnamefont {N.~V.}\ \bibnamefont {{Nitta}}},
  \bibinfo {author} {\bibfnamefont {C.~J.}\ \bibnamefont {{Schrijver}}},
  \bibinfo {author} {\bibfnamefont {A.~M.}\ \bibnamefont {{Title}}}, \ and\
  \bibinfo {author} {\bibfnamefont {T.~D.}\ \bibnamefont {{Tarbell}}},\ }\href  {\doibase 10.1088/2041-8205/723/1/L53} {\bibfield  {journal} {\bibinfo
  {journal} {\apjl}\ }\textbf {\bibinfo {volume} {723}},\ \unskip\ \bibinfo
  {pages} {L53--L59}, November (\bibinfo {year} {2010})}\BibitemShut {NoStop}%
\bibitem [{\citenamefont {{Liu}}\ \emph {et~al.}(2011)\citenamefont {{Liu}},
  \citenamefont {{Title}}, \citenamefont {{Zhao}}, \citenamefont {{Ofman}},
  \citenamefont {{Schrijver}}, \citenamefont {{Aschwanden}}, \citenamefont {{De
  Pontieu}},\ and\ \citenamefont
  {{Tarbell}}}]{LiuW.FastWave.2011ApJ...736L..13L}%
  \BibitemOpen
  \bibfield  {author} {\bibinfo {author} {\bibfnamefont {W.}~\bibnamefont
  {{Liu}}}, \bibinfo {author} {\bibfnamefont {A.~M.}\ \bibnamefont {{Title}}},
  \bibinfo {author} {\bibfnamefont {J.}~\bibnamefont {{Zhao}}}, \bibinfo
  {author} {\bibfnamefont {L.}~\bibnamefont {{Ofman}}}, \ and\
\bibinfo {author} {\bibfnamefont {C.~J.}\ \bibnamefont {{Schrijver}}}, 
 \bibinfo {author} {\bibfnamefont {et}\ \bibnamefont {{al.}}},\ }
\href {\doibase 10.1088/2041-8205/736/1/L13} {\bibfield  {journal} {\bibinfo  {journal}
  {\apjl}\ }\textbf {\bibinfo {volume} {736}},\ p.\ \bibinfo {pages} {L13}, July
  (\bibinfo {year} {2011})}. 	
\bibitem [{\citenamefont {{Liu}}\ \emph {et~al.}(2012)\citenamefont {{Liu}},
  \citenamefont {{Ofman}}, \citenamefont {{Nitta}}, \citenamefont
  {{Aschwanden}}, \citenamefont {{Schrijver}}, \citenamefont {{Title}},\ and\
  \citenamefont {{Tarbell}}}]{LiuW.cavity-oscil.2012ApJ...753...52L}%
  \BibitemOpen
  \bibfield  {author} {\bibinfo {author} {\bibfnamefont {W.}~\bibnamefont
  {{Liu}}}, \bibinfo {author} {\bibfnamefont {L.}~\bibnamefont {{Ofman}}},
 \bibinfo {author} {\bibfnamefont {N.~V.}\ \bibnamefont {{Nitta}}}, 
 \bibinfo {author} {\bibfnamefont {M.~J.}\ \bibnamefont {{Aschwanden}}},
\ and\ \bibinfo  {author} {\bibfnamefont {C.~J.}\ \bibnamefont {{Schrijver}}}, 
 \bibinfo {author} {\bibfnamefont {et}\ \bibnamefont {{al.}}},\ }
\href {\doibase  10.1088/0004-637X/753/1/52} {\bibfield  {journal} {\bibinfo  {journal}
  {\apj}\ }\textbf {\bibinfo {volume} {753}},\ p.~\bibinfo {pages} {52}, July
  (\bibinfo {year} {2012})}.
\bibitem [{\citenamefont {{Ofman}}\ \emph {et~al.}(2011)\citenamefont
  {{Ofman}}, \citenamefont {{Liu}}, \citenamefont {{Title}},\ and\
  \citenamefont {{Aschwanden}}}]{Ofman.Liu.fast-wave.2011ApJ...740L..33O}%
  \BibitemOpen
  \bibfield  {author} {\bibinfo {author} {\bibfnamefont {L.}~\bibnamefont
  {{Ofman}}}, \bibinfo {author} {\bibfnamefont {W.}~\bibnamefont {{Liu}}},
  \bibinfo {author} {\bibfnamefont {A.}~\bibnamefont {{Title}}}, \ and\
  \bibinfo {author} {\bibfnamefont {M.}~\bibnamefont {{Aschwanden}}},\ }
\href  {\doibase 10.1088/2041-8205/740/2/L33} {\bibfield  {journal} {\bibinfo
  {journal} {\apjl}\ }\textbf {\bibinfo {volume} {740}},\ p.\ \bibinfo {pages}
  {L33}, October (\bibinfo {year} {2011})}\BibitemShut {NoStop}%
\bibitem [{\citenamefont {{Pascoe}}, \citenamefont {{Nakariakov}},\ and\
  \citenamefont
  {{Kupriyanova}}(2013)}]{Pascoe.wing.QFPs.funnel.2D.MHD.2013A&A...560A..97P}%
  \BibitemOpen
  \bibfield  {author} {\bibinfo {author} {\bibfnamefont {D.~J.}\ \bibnamefont
  {{Pascoe}}}, \bibinfo {author} {\bibfnamefont {V.~M.}\ \bibnamefont
  {{Nakariakov}}}, \ and\ \bibinfo {author} {\bibfnamefont {E.~G.}\
  \bibnamefont {{Kupriyanova}}},\ }\href {\doibase 10.1051/0004-6361/201322678}
  {\bibfield  {journal} {\bibinfo  {journal} {\aap}\ }\textbf {\bibinfo
  {volume} {560}},\ p.\ \bibinfo {pages} {A97}, December (\bibinfo {year}
  {2013})}\BibitemShut {NoStop}%
\bibitem [{\citenamefont {{Pascoe}}, \citenamefont {{Nakariakov}},\ and\
  \citenamefont
  {{Kupriyanova}}(2014)}]{PascoeDavid.coronal.hole.anti-wave-guide.QFPs.2014A&%
A...568A..20P}%
  \BibitemOpen
  \bibfield  {author} {\bibinfo {author} {\bibfnamefont {D.~J.}\ \bibnamefont
  {{Pascoe}}}, \bibinfo {author} {\bibfnamefont {V.~M.}\ \bibnamefont
  {{Nakariakov}}}, \ and\ \bibinfo {author} {\bibfnamefont {E.~G.}\
  \bibnamefont {{Kupriyanova}}},\ }\href {\doibase 10.1051/0004-6361/201423931}
  {\bibfield  {journal} {\bibinfo  {journal} {\aap}\ }\textbf {\bibinfo
  {volume} {568}},\ p.\ \bibinfo {pages} {A20}, August (\bibinfo {year}
  {2014})}\BibitemShut {NoStop}%
\bibitem [{\citenamefont {{Yang}}\ \emph {et~al.}(2015)\citenamefont {{Yang}},
  \citenamefont {{Zhang}}, \citenamefont {{He}}, \citenamefont {{Peter}},
  \citenamefont {{Tu}}, \citenamefont {{Wang}}, \citenamefont {{Zhang}},\ and\
  \citenamefont {{Feng}}}]{YangLiping.QFPs.plasmoid.simul.2015ApJ...800..111Y}%
  \BibitemOpen
  \bibfield  {author} {\bibinfo {author} {\bibfnamefont {L.}~\bibnamefont
  {{Yang}}}, \bibinfo {author} {\bibfnamefont {L.}~\bibnamefont {{Zhang}}},
  \bibinfo {author} {\bibfnamefont {J.}~\bibnamefont {{He}}}, \bibinfo {author}
  {\bibfnamefont {H.}~\bibnamefont {{Peter}}}, \bibinfo {author} {\bibfnamefont
  {C.}~\bibnamefont {{Tu}}},
\ and\ \bibinfo {author} {\bibfnamefont {L.}~\bibnamefont {{Wang}}}, 
 \bibinfo {author} {\bibfnamefont {et}\ \bibnamefont {{al.}}},\ }
\href  {\doibase 10.1088/0004-637X/800/2/111} {\bibfield  {journal} {\bibinfo
  {journal} {\apj}\ }\textbf {\bibinfo {volume} {800}},\ p.\ \bibinfo {pages}
  {111}, February (\bibinfo {year} {2015})}\BibitemShut {NoStop}%
\bibitem [{\citenamefont {{Young}}\ \emph {et~al.}(1961)\citenamefont
  {{Young}}, \citenamefont {{Spencer}}, \citenamefont {{Moreton}},\ and\
  \citenamefont {{Roberts}}}]{YoungCW.flare.radio.QPP.1961ApJ...133..243Y}%
  \BibitemOpen
  \bibfield  {author} {\bibinfo {author} {\bibfnamefont {C.~W.}\ \bibnamefont
  {{Young}}}, \bibinfo {author} {\bibfnamefont {C.~L.}\ \bibnamefont
  {{Spencer}}}, \bibinfo {author} {\bibfnamefont {G.~E.}\ \bibnamefont
  {{Moreton}}}, \ and\ \bibinfo {author} {\bibfnamefont {J.~A.}\ \bibnamefont
  {{Roberts}}},\ }\href {\doibase 10.1086/147019} {\bibfield  {journal}
  {\bibinfo  {journal} {\apj}\ }\textbf {\bibinfo {volume} {133}},\ p.\
  \bibinfo {pages} {243}, January (\bibinfo {year} {1961})}\BibitemShut {NoStop}%
\bibitem [{\citenamefont {{Nakariakov}}\ and\ \citenamefont
  {{Melnikov}}(2009)}]{Nakariakov.Melnikov.QPP.2009SSRv..149..119N}%
  \BibitemOpen
  \bibfield  {author} {\bibinfo {author} {\bibfnamefont {V.~M.}\ \bibnamefont
  {{Nakariakov}}}\ and\ \bibinfo {author} {\bibfnamefont {V.~F.}\ \bibnamefont
  {{Melnikov}}},\ }\href {\doibase 10.1007/s11214-009-9536-3} {\bibfield
  {journal} {\bibinfo  {journal} {\ssr}\ }\textbf {\bibinfo {volume} {149}},\
  \unskip\ \bibinfo {pages} {119--151}, December (\bibinfo {year}
  {2009})}\BibitemShut {NoStop}%
\bibitem [{\citenamefont
  {{Uchida}}(1970)}]{Uchida.coronal-seismology.1970PASJ...22..341U}%
  \BibitemOpen
  \bibfield  {author} {\bibinfo {author} {\bibfnamefont {Y.}~\bibnamefont
  {{Uchida}}},\ }\href@noop {} {\bibfield  {journal} {\bibinfo  {journal}
  {\pasj}\ }\textbf {\bibinfo {volume} {22}},\ p.\ \bibinfo {pages} {341}
  (\bibinfo {year} {1970})}\BibitemShut {NoStop}%
\bibitem [{\citenamefont {{Roberts}}, \citenamefont {{Edwin}},\ and\
  \citenamefont
  {{Benz}}(1984)}]{Roberts.coronal-seismology.1984ApJ...279..857R}%
  \BibitemOpen
  \bibfield  {author} {\bibinfo {author} {\bibfnamefont {B.}~\bibnamefont
  {{Roberts}}}, \bibinfo {author} {\bibfnamefont {P.~M.}\ \bibnamefont
  {{Edwin}}}, \ and\ \bibinfo {author} {\bibfnamefont {A.~O.}\ \bibnamefont
  {{Benz}}},\ }\href {\doibase 10.1086/161956} {\bibfield  {journal} {\bibinfo
  {journal} {\apj}\ }\textbf {\bibinfo {volume} {279}},\ \unskip\ \bibinfo
  {pages} {857--865}, April (\bibinfo {year} {1984})}\BibitemShut {NoStop}%
\bibitem [{\citenamefont {{Shen}}\ and\ \citenamefont
  {{Liu}}(2012)}]{ShenYD.LiuY.QFP.wave.2012ApJ...753...53S}%
  \BibitemOpen
  \bibfield  {author} {\bibinfo {author} {\bibfnamefont {Y.}~\bibnamefont
  {{Shen}}}\ and\ \bibinfo {author} {\bibfnamefont {Y.}~\bibnamefont {{Liu}}},\
  }\href {\doibase 10.1088/0004-637X/753/1/53} {\bibfield  {journal} {\bibinfo
  {journal} {\apj}\ }\textbf {\bibinfo {volume} {753}},\ p.~\bibinfo {pages}
  {53}, July (\bibinfo {year} {2012})}.
\bibitem [{\citenamefont {{Shen}}\ \emph {et~al.}(2013)\citenamefont {{Shen}},
  \citenamefont {{Liu}}, \citenamefont {{Su}}, \citenamefont {{Li}},
  \citenamefont {{Zhang}}, \citenamefont {{Tian}}, \citenamefont {{Zhao}},\
  and\ \citenamefont
  {{Elmhamdi}}}]{ShenYD.LiuYu.QFP.171.193.2013SoPh..288..585S}%
  \BibitemOpen
  \bibfield  {author} {\bibinfo {author} {\bibfnamefont {Y.-D.}\ \bibnamefont
  {{Shen}}}, \bibinfo {author} {\bibfnamefont {Y.}~\bibnamefont {{Liu}}},
  \bibinfo {author} {\bibfnamefont {J.-T.}\ \bibnamefont {{Su}}}, \bibinfo
  {author} {\bibfnamefont {H.}~\bibnamefont {{Li}}}, 
\ and\  \bibinfo {author} {\bibfnamefont {X.-F.}\ \bibnamefont {{Zhang}}}, 
 \bibinfo {author} {\bibfnamefont {et}\ \bibnamefont {{al.}}},\ }
\href {\doibase  10.1007/s11207-013-0395-4} {\bibfield  {journal} {\bibinfo  {journal}
  {\solphys}\ }\textbf {\bibinfo {volume} {288}},\ \unskip\ \bibinfo {pages}
  {585--602}, December (\bibinfo {year} {2013})}.
\bibitem [{\citenamefont {{Yuan}}\ \emph {et~al.}(2013)\citenamefont {{Yuan}},
  \citenamefont {{Shen}}, \citenamefont {{Liu}}, \citenamefont {{Nakariakov}},
  \citenamefont {{Tan}},\ and\ \citenamefont
  {{Huang}}}]{YuanD.QFP.distinct.trains.2013A&A...554A.144Y}%
  \BibitemOpen
  \bibfield  {author} {\bibinfo {author} {\bibfnamefont {D.}~\bibnamefont
  {{Yuan}}}, \bibinfo {author} {\bibfnamefont {Y.}~\bibnamefont {{Shen}}},
  \bibinfo {author} {\bibfnamefont {Y.}~\bibnamefont {{Liu}}}, \bibinfo
  {author} {\bibfnamefont {V.~M.}\ \bibnamefont {{Nakariakov}}}, \bibinfo
  {author} {\bibfnamefont {B.}~\bibnamefont {{Tan}}}, \ and\ \bibinfo {author}
  {\bibfnamefont {J.}~\bibnamefont {{Huang}}},\ }\href {\doibase
  10.1051/0004-6361/201321435} {\bibfield  {journal} {\bibinfo  {journal}
  {\aap}\ }\textbf {\bibinfo {volume} {554}},\ p.\ \bibinfo {pages} {A144}, June
  (\bibinfo {year} {2013})}\BibitemShut {NoStop}%
\bibitem [{\citenamefont {{Nistic{\`o}}}, \citenamefont {{Pascoe}},\ and\
  \citenamefont
  {{Nakariakov}}(2014)}]{NisticoG.2013Dec7.M1.2.QFP.2014AA...569A..12N}%
  \BibitemOpen
  \bibfield  {author} {\bibinfo {author} {\bibfnamefont {G.}~\bibnamefont
  {{Nistic{\`o}}}}, \bibinfo {author} {\bibfnamefont {D.~J.}\ \bibnamefont
  {{Pascoe}}}, \ and\ \bibinfo {author} {\bibfnamefont {V.~M.}\ \bibnamefont
  {{Nakariakov}}},\ }\href {\doibase 10.1051/0004-6361/201423763} {\bibfield
  {journal} {\bibinfo  {journal} {\aap}\ }\textbf {\bibinfo {volume} {569}},\
  p.\ \bibinfo {pages} {A12}, September (\bibinfo {year} {2014})}\BibitemShut
  {NoStop}%
\bibitem [{\citenamefont {{Zhang}}\ \emph {et~al.}(2015)\citenamefont
  {{Zhang}}, \citenamefont {{Zhang}}, \citenamefont {{Wang}},\ and\
  \citenamefont
  {{Nakariakov}}}]{ZhangYuZong.fast-slow-modes.2015A&A...581A..78Z}%
  \BibitemOpen
  \bibfield  {author} {\bibinfo {author} {\bibfnamefont {Y.}~\bibnamefont
  {{Zhang}}}, \bibinfo {author} {\bibfnamefont {J.}~\bibnamefont {{Zhang}}},
  \bibinfo {author} {\bibfnamefont {J.}~\bibnamefont {{Wang}}}, \ and\ \bibinfo
  {author} {\bibfnamefont {V.~M.}\ \bibnamefont {{Nakariakov}}},\ }
\href  {\doibase 10.1051/0004-6361/201525621} {\bibfield  {journal} {\bibinfo
  {journal} {\aap}\ }\textbf {\bibinfo {volume} {581}},\ p.\ \bibinfo {pages}
  {A78}, September (\bibinfo {year} {2015})}\BibitemShut {NoStop}%
\bibitem [{\citenamefont {{Liu}}\ and\ \citenamefont
  {{Ofman}}(2014)}]{LiuW.OfmanL.EUV.wave.review.2014SoPh..289.3233L}%
  \BibitemOpen
  \bibfield  {author} {\bibinfo {author} {\bibfnamefont {W.}~\bibnamefont
  {{Liu}}}\ and\ \bibinfo {author} {\bibfnamefont {L.}~\bibnamefont
  {{Ofman}}},\ }\href {\doibase 10.1007/s11207-014-0528-4} {\bibfield
  {journal} {\bibinfo  {journal} {\solphys}\ }\textbf {\bibinfo {volume}
  {289}},\ \unskip\ \bibinfo {pages} {3233--3277}, September (\bibinfo {year}
  {2014})}.
\bibitem [{\citenamefont {{Schrijver}}\ and\ \citenamefont
  {{Title}}(2011)}]{Schrijver.Title.2010Aug01-long-range-couple.2011JGRA..116.%
4108S}%
  \BibitemOpen
  \bibfield  {author} {\bibinfo {author} {\bibfnamefont {C.~J.}\ \bibnamefont
  {{Schrijver}}}\ and\ \bibinfo {author} {\bibfnamefont {A.~M.}\ \bibnamefont
  {{Title}}},\ }\href {\doibase 10.1029/2010JA016224} {\bibfield  {journal}
  {\bibinfo  {journal} {\jgrsp}\ }\textbf {\bibinfo {volume} {116}},\ p.\
  \bibinfo {pages} {4108} April (\bibinfo {year} {2011})}\BibitemShut {NoStop}%
\bibitem [{\citenamefont
  {{Karlick{\'y}}}(2014)}]{KarlickyM.zebra.my-2010Aug01-QFP.2014AA...561A..34K%
}%
  \BibitemOpen
  \bibfield  {author} {\bibinfo {author} {\bibfnamefont {M.}~\bibnamefont
  {{Karlick{\'y}}}},\ }\href {\doibase 10.1051/0004-6361/201322547} {\bibfield
  {journal} {\bibinfo  {journal} {\aap}\ }\textbf {\bibinfo {volume} {561}},\
  p.\ \bibinfo {pages} {A34}, January (\bibinfo {year} {2014})}\BibitemShut
  {NoStop}%
\bibitem [{\citenamefont {{Tomczyk}}\ and\ \citenamefont
  {{McIntosh}}(2009)}]{TomczykMcIntosh.coronal-time-distance-seism.2009ApJ...6%
97.1384T}%
  \BibitemOpen
  \bibfield  {author} {\bibinfo {author} {\bibfnamefont {S.}~\bibnamefont
  {{Tomczyk}}}\ and\ \bibinfo {author} {\bibfnamefont {S.~W.}\ \bibnamefont
  {{McIntosh}}},\ }\href {\doibase 10.1088/0004-637X/697/2/1384} {\bibfield
  {journal} {\bibinfo  {journal} {\apj}\ }\textbf {\bibinfo {volume} {697}},\
  \unskip\ \bibinfo {pages} {1384--1391}, June (\bibinfo {year} {2009})}.
\bibitem [{\citenamefont {{Ireland}}, \citenamefont {{McAteer}},\ and\
  \citenamefont
  {{Inglis}}(2015)}]{IrelandJ.corona.power.spec.2015ApJ...798....1I}%
  \BibitemOpen
  \bibfield  {author} {\bibinfo {author} {\bibfnamefont {J.}~\bibnamefont
  {{Ireland}}}, \bibinfo {author} {\bibfnamefont {R.~T.~J.}\ \bibnamefont
  {{McAteer}}}, \ and\ \bibinfo {author} {\bibfnamefont {A.~R.}\ \bibnamefont
  {{Inglis}}},\ }\href {\doibase 10.1088/0004-637X/798/1/1} {\bibfield
  {journal} {\bibinfo  {journal} {\apj}\ }\textbf {\bibinfo {volume} {798}},\
  p.~\bibinfo {pages} {1}, January (\bibinfo {year} {2015})}.
\bibitem [{\citenamefont {{Karlick{\'y}}}, \citenamefont
  {{M{\'e}sz{\'a}rosov{\'a}}},\ and\ \citenamefont
  {{Jel{\'{\i}}nek}}(2013)}]{KarlickyM.fiber.burst.fast.sausage.wave.train.201%
3A&A...550A...1K}%
  \BibitemOpen
  \bibfield  {author} {\bibinfo {author} {\bibfnamefont {M.}~\bibnamefont
  {{Karlick{\'y}}}}, \bibinfo {author} {\bibfnamefont {H.}~\bibnamefont
  {{M{\'e}sz{\'a}rosov{\'a}}}}, \ and\ \bibinfo {author} {\bibfnamefont
  {P.}~\bibnamefont {{Jel{\'{\i}}nek}}},\ }\href {\doibase
  10.1051/0004-6361/201220296} {\bibfield  {journal} {\bibinfo  {journal}
  {\aap}\ }\textbf {\bibinfo {volume} {550}},\ p.~\bibinfo {pages} {A1}, February
  (\bibinfo {year} {2013})}.
\bibitem [{\citenamefont
  {{Karlick{\'y}}}(2013)}]{KarlickyMarian.zebra.mudolated.by.fastMode.2013A&A.%
..552A..90K}%
  \BibitemOpen
  \bibfield  {author} {\bibinfo {author} {\bibfnamefont {M.}~\bibnamefont
  {{Karlick{\'y}}}},\ }\href {\doibase 10.1051/0004-6361/201321356} {\bibfield
  {journal} {\bibinfo  {journal} {\aap}\ }\textbf {\bibinfo {volume} {552}},\
  p.\ \bibinfo {pages} {A90}, April (\bibinfo {year} {2013})}\BibitemShut
  {NoStop}%
\bibitem [{\citenamefont
  {{Aschwanden}}(2002)}]{AschwandenM2002SSRv..101....1A}%
  \BibitemOpen
  \bibfield  {author} {\bibinfo {author} {\bibfnamefont {M.~J.}\ \bibnamefont
  {{Aschwanden}}},\ }\href {\doibase 10.1023/A:1019712124366} {\bibfield
  {journal} {\bibinfo  {journal} {\ssr}\ }\textbf {\bibinfo {volume} {101}},\
  \unskip\ \bibinfo {pages} {1--227}, January (\bibinfo {year}
  {2002})}\BibitemShut {NoStop}%
\bibitem [{\citenamefont {{Downs}}\ \emph {et~al.}(2012)\citenamefont
  {{Downs}}, \citenamefont {{Roussev}}, \citenamefont {{van der Holst}},
  \citenamefont {{Lugaz}},\ and\ \citenamefont
  {{Sokolov}}}]{DownsC.MHD.2010-06-13-AIA-wave.2012ApJ...750..134D}%
  \BibitemOpen
  \bibfield  {author} {\bibinfo {author} {\bibfnamefont {C.}~\bibnamefont
  {{Downs}}}, \bibinfo {author} {\bibfnamefont {I.~I.}\ \bibnamefont
  {{Roussev}}}, \bibinfo {author} {\bibfnamefont {B.}~\bibnamefont {{van der
  Holst}}}, \bibinfo {author} {\bibfnamefont {N.}~\bibnamefont {{Lugaz}}}, \
  and\ \bibinfo {author} {\bibfnamefont {I.~V.}\ \bibnamefont {{Sokolov}}},\
  }\href {\doibase 10.1088/0004-637X/750/2/134} {\bibfield  {journal} {\bibinfo
   {journal} {\apj}\ }\textbf {\bibinfo {volume} {750}},\ p.\ \bibinfo {pages}
  {134}, May (\bibinfo {year} {2012})}\BibitemShut {NoStop}%
\bibitem [{\citenamefont {{Su}}\ \emph {et~al.}(2011)\citenamefont {{Su}},
  \citenamefont {{Surges}}, \citenamefont {{van Ballegooijen}}, \citenamefont
  {{DeLuca}},\ and\ \citenamefont
  {{Golub}}}]{SuYN.flux.rope.insert.CME.2010Apr08.2011ApJ...734...53S}%
  \BibitemOpen
  \bibfield  {author} {\bibinfo {author} {\bibfnamefont {Y.}~\bibnamefont
  {{Su}}}, \bibinfo {author} {\bibfnamefont {V.}~\bibnamefont {{Surges}}},
  \bibinfo {author} {\bibfnamefont {A.}~\bibnamefont {{van Ballegooijen}}},
  \bibinfo {author} {\bibfnamefont {E.}~\bibnamefont {{DeLuca}}}, \ and\
  \bibinfo {author} {\bibfnamefont {L.}~\bibnamefont {{Golub}}},\ }
\href  {\doibase 10.1088/0004-637X/734/1/53} {\bibfield  {journal} {\bibinfo
  {journal} {\apj}\ }\textbf {\bibinfo {volume} {734}},\ p.~\bibinfo {pages}
  {53}, June (\bibinfo {year} {2011})}\BibitemShut {NoStop}%
\bibitem [{\citenamefont {{Ofman}}\ and\ \citenamefont
  {{Thompson}}(2011)}]{Ofman.Thompson.AIA.KH.instab.2011ApJ...734L..11O}%
  \BibitemOpen
  \bibfield  {author} {\bibinfo {author} {\bibfnamefont {L.}~\bibnamefont
  {{Ofman}}}\ and\ \bibinfo {author} {\bibfnamefont {B.~J.}\ \bibnamefont
  {{Thompson}}},\ }\href {\doibase 10.1088/2041-8205/734/1/L11} {\bibfield
  {journal} {\bibinfo  {journal} {\apjl}\ }\textbf {\bibinfo {volume} {734}},\
  p.\ \bibinfo {pages} {L11}, June (\bibinfo {year} {2011})}.
\bibitem [{\citenamefont {{Nitta}}\ \emph {et~al.}(2013)\citenamefont
  {{Nitta}}, \citenamefont {{Schrijver}}, \citenamefont {{Title}},\ and\
  \citenamefont {{Liu}}}]{NittaN.AIA.wave.stat.2013ApJ...776...58N}%
  \BibitemOpen
  \bibfield  {author} {\bibinfo {author} {\bibfnamefont {N.~V.}\ \bibnamefont
  {{Nitta}}}, \bibinfo {author} {\bibfnamefont {C.~J.}\ \bibnamefont
  {{Schrijver}}}, \bibinfo {author} {\bibfnamefont {A.~M.}\ \bibnamefont
  {{Title}}}, \ and\ \bibinfo {author} {\bibfnamefont {W.}~\bibnamefont
  {{Liu}}},\ }\href {\doibase 10.1088/0004-637X/776/1/58} {\bibfield  {journal}
  {\bibinfo  {journal} {\apj}\ }\textbf {\bibinfo {volume} {776}},\ p.~\bibinfo
  {pages} {58}, October (\bibinfo {year} {2013})}.
\bibitem [{\citenamefont {{Sun}}\ \emph {et~al.}(2015)\citenamefont {{Sun}},
  \citenamefont {{Bobra}}, \citenamefont {{Hoeksema}}, \citenamefont {{Liu}},
  \citenamefont {{Li}}, \citenamefont {{Shen}}, \citenamefont {{Couvidat}},
  \citenamefont {{Norton}},\ and\ \citenamefont
  {{Fisher}}}]{SunXudong.AR12192.flareBig.CMEpoor.2015ApJ...804L..28S}%
  \BibitemOpen
  \bibfield  {author} {\bibinfo {author} {\bibfnamefont {X.}~\bibnamefont
  {{Sun}}}, \bibinfo {author} {\bibfnamefont {M.~G.}\ \bibnamefont {{Bobra}}},
  \bibinfo {author} {\bibfnamefont {J.~T.}\ \bibnamefont {{Hoeksema}}},
  \bibinfo {author} {\bibfnamefont {Y.}~\bibnamefont {{Liu}}}, \bibinfo
  {author} {\bibfnamefont {Y.}~\bibnamefont {{Li}}}, \bibinfo {author}
\ and\  {\bibfnamefont {C.}~\bibnamefont {{Shen}}},
\bibinfo {author} {\bibfnamefont {et}\ \bibnamefont {{al.}}},\ }
\href {\doibase 10.1088/2041-8205/804/2/L28}
  {\bibfield  {journal} {\bibinfo  {journal} {\apjl}\ }\textbf {\bibinfo
  {volume} {804}},\ p.\ \bibinfo {pages} {L28}, May (\bibinfo {year} {2015})}.
\bibitem [{\citenamefont {{Chen}}\ \emph {et~al.}(2015)\citenamefont {{Chen}},
  \citenamefont {{Zhang}}, \citenamefont {{Ma}}, \citenamefont {{Yang}},
  \citenamefont {{Li}}, \citenamefont {{Huang}},\ and\ \citenamefont
  {{Xiao}}}]{ChenHuadong.AR12192.confined.flares.2015ApJ...808L..24C}%
  \BibitemOpen
  \bibfield  {author} {\bibinfo {author} {\bibfnamefont {H.}~\bibnamefont
  {{Chen}}}, \bibinfo {author} {\bibfnamefont {J.}~\bibnamefont {{Zhang}}},
  \bibinfo {author} {\bibfnamefont {S.}~\bibnamefont {{Ma}}}, \bibinfo {author}
  {\bibfnamefont {S.}~\bibnamefont {{Yang}}}, \bibinfo {author} {\bibfnamefont
  {L.}~\bibnamefont {{Li}}}, \bibinfo {author} {\bibfnamefont {X.}~\bibnamefont
  {{Huang}}}, \ and\ \bibinfo {author} {\bibfnamefont {J.}~\bibnamefont
  {{Xiao}}},\ }\href {\doibase 10.1088/2041-8205/808/1/L24} {\bibfield
  {journal} {\bibinfo  {journal} {\apjl}\ }\textbf {\bibinfo {volume} {808}},\
  p.\ \bibinfo {pages} {L24}, July (\bibinfo {year} {2015})}.
\bibitem [{\citenamefont {{Habbal}}\ \emph {et~al.}(2011)\citenamefont
  {{Habbal}}, \citenamefont {{Druckm{\"u}ller}}, \citenamefont {{Morgan}},
  \citenamefont {{Ding}}, \citenamefont {{Johnson}}, \citenamefont
  {{Druckm{\"u}llerov{\'a}}}, \citenamefont {{Daw}}, \citenamefont {{Arndt}},
  \citenamefont {{Dietzel}},\ and\ \citenamefont
  {{Saken}}}]{HabbalS.cone.wake.2011ApJ...734..120H}%
  \BibitemOpen
  \bibfield  {author} {\bibinfo {author} {\bibfnamefont {S.~R.}\ \bibnamefont
  {{Habbal}}}, \bibinfo {author} {\bibfnamefont {M.}~\bibnamefont
  {{Druckm{\"u}ller}}}, \bibinfo {author} {\bibfnamefont {H.}~\bibnamefont
  {{Morgan}}}, \bibinfo {author} {\bibfnamefont {A.}~\bibnamefont {{Ding}}},
\ and\  \bibinfo {author} {\bibfnamefont {J.}~\bibnamefont {{Johnson}}}, 
\bibinfo {author} {\bibfnamefont {et}\ \bibnamefont {{al.}}},\ }
\href {\doibase 10.1088/0004-637X/734/2/120} {\bibfield  {journal} {\bibinfo  {journal}
  {\apj}\ }\textbf {\bibinfo {volume} {734}},\ p.\ \bibinfo {pages} {120}, June
  (\bibinfo {year} {2011})}\BibitemShut {NoStop}%
\bibitem [{\citenamefont {{Alazraki}}\ and\ \citenamefont
  {{Couturier}}(1971)}]{AlazrakiG.Alfven.acc.solar.wind.1971A&A....13..380A}%
  \BibitemOpen
  \bibfield  {author} {\bibinfo {author} {\bibfnamefont {G.}~\bibnamefont
  {{Alazraki}}}\ and\ \bibinfo {author} {\bibfnamefont {P.}~\bibnamefont
  {{Couturier}}},\ }\href@noop {} {\bibfield  {journal} {\bibinfo  {journal}
  {\aap}\ }\textbf {\bibinfo {volume} {13}},\ p.\ \bibinfo {pages} {380}, August
  (\bibinfo {year} {1971})}\BibitemShut {NoStop}%
\bibitem [{\citenamefont {{Fla}}\ \emph {et~al.}(1984)\citenamefont {{Fla}},
  \citenamefont {{Habbal}}, \citenamefont {{Holzer}},\ and\ \citenamefont
  {{Leer}}}]{FlaT.fast.mode.CH.solar.wind.1984ApJ...280..382F}%
  \BibitemOpen
  \bibfield  {author} {\bibinfo {author} {\bibfnamefont {T.}~\bibnamefont
  {{Fla}}}, \bibinfo {author} {\bibfnamefont {S.~R.}\ \bibnamefont {{Habbal}}},
  \bibinfo {author} {\bibfnamefont {T.~E.}\ \bibnamefont {{Holzer}}}, \ and\
  \bibinfo {author} {\bibfnamefont {E.}~\bibnamefont {{Leer}}},\ }
\href  {\doibase 10.1086/162003} {\bibfield  {journal} {\bibinfo  {journal} {\apj}\
  }\textbf {\bibinfo {volume} {280}},\ \unskip\ \bibinfo {pages} {382--390}, May
  (\bibinfo {year} {1984})}\BibitemShut {NoStop}%
\bibitem [{\citenamefont
  {{Parker}}(1991)}]{ParkerE.phase.mixing.CH.SW.1991ApJ...376..355P}%
  \BibitemOpen
  \bibfield  {author} {\bibinfo {author} {\bibfnamefont {E.~N.}\ \bibnamefont
  {{Parker}}},\ }\href {\doibase 10.1086/170285} {\bibfield  {journal}
  {\bibinfo  {journal} {\apj}\ }\textbf {\bibinfo {volume} {376}},\ \unskip\
  \bibinfo {pages} {355--363}, July (\bibinfo {year} {1991})}\BibitemShut
  {NoStop}%
\bibitem [{\citenamefont
  {{Ofman}}(2010)}]{Ofman.solarwind-review.2010LRSP....7....4O}%
  \BibitemOpen
  \bibfield  {author} {\bibinfo {author} {\bibfnamefont {L.}~\bibnamefont
  {{Ofman}}},\ }\href {\doibase 10.12942/lrsp-2010-4} {\bibfield  {journal}
  {\bibinfo  {journal} {\lrsp}\ }\textbf {\bibinfo {volume} {7}},\ p.~\bibinfo
  {pages} {4}, October (\bibinfo {year} {2010})}\BibitemShut {NoStop}%
\bibitem [{\citenamefont {{Liu}}, \citenamefont {{Chen}},\ and\ \citenamefont
  {{Petrosian}}(2013)}]{LiuW.cusp.flare.20120719M7.2013ApJ...767..168L}%
  \BibitemOpen
  \bibfield  {author} {\bibinfo {author} {\bibfnamefont {W.}~\bibnamefont
  {{Liu}}}, \bibinfo {author} {\bibfnamefont {Q.}~\bibnamefont {{Chen}}}, \
  and\ \bibinfo {author} {\bibfnamefont {V.}~\bibnamefont {{Petrosian}}},\
  }\href {\doibase 10.1088/0004-637X/767/2/168} {\bibfield  {journal} {\bibinfo
   {journal} {\apj}\ }\textbf {\bibinfo {volume} {767}},\ p.\ \bibinfo {pages}
  {168}, April (\bibinfo {year} {2013})}.
\bibitem [{\citenamefont {{Ofman}}\ and\ \citenamefont
  {{Sui}}(2006)}]{OfmanSui.HXR-oscil.2006ApJ...644L.149O}%
  \BibitemOpen
  \bibfield  {author} {\bibinfo {author} {\bibfnamefont {L.}~\bibnamefont
  {{Ofman}}}\ and\ \bibinfo {author} {\bibfnamefont {L.}~\bibnamefont
  {{Sui}}},\ }\href {\doibase 10.1086/505622} {\bibfield  {journal} {\bibinfo
  {journal} {\apjl}\ }\textbf {\bibinfo {volume} {644}},\ \unskip\ \bibinfo
  {pages} {L149--L152}, June (\bibinfo {year} {2006})}\BibitemShut {NoStop}%
\bibitem [{\citenamefont {{De Moortel}}\ \emph {et~al.}(2002)\citenamefont {{De
  Moortel}}, \citenamefont {{Ireland}}, \citenamefont {{Hood}},\ and\
  \citenamefont
  {{Walsh}}}]{DeMoortelI.3-5min.slow-mode.leak.2corona.2002A&A...387L..13D}%
  \BibitemOpen
  \bibfield  {author} {\bibinfo {author} {\bibfnamefont {I.}~\bibnamefont {{De
  Moortel}}}, \bibinfo {author} {\bibfnamefont {J.}~\bibnamefont {{Ireland}}},
  \bibinfo {author} {\bibfnamefont {A.~W.}\ \bibnamefont {{Hood}}}, \ and\
  \bibinfo {author} {\bibfnamefont {R.~W.}\ \bibnamefont {{Walsh}}},\ }
\href  {\doibase 10.1051/0004-6361:20020436} {\bibfield  {journal} {\bibinfo
  {journal} {\aap}\ }\textbf {\bibinfo {volume} {387}},\ \unskip\ \bibinfo
  {pages} {L13--L16}, May (\bibinfo {year} {2002})}\BibitemShut {NoStop}%
\bibitem [{\citenamefont {{Bogdan}}\ \emph {et~al.}(2003)\citenamefont
  {{Bogdan}}, \citenamefont {{Carlsson}}, \citenamefont {{Hansteen}},
  \citenamefont {{McMurry}}, \citenamefont {{Rosenthal}}, \citenamefont
  {{Johnson}}, \citenamefont {{Petty-Powell}}, \citenamefont {{Zita}},
  \citenamefont {{Stein}}, \citenamefont {{McIntosh}},\ and\ \citenamefont
  {{Nordlund}}}]{Bogdan.coronal-wave-simul.2003ApJ...599..626B}%
  \BibitemOpen
  \bibfield  {author} {\bibinfo {author} {\bibfnamefont {T.~J.}\ \bibnamefont
  {{Bogdan}}}, \bibinfo {author} {\bibfnamefont {M.}~\bibnamefont
  {{Carlsson}}}, \bibinfo {author} {\bibfnamefont {V.~H.}\ \bibnamefont {{Hansteen}}}, 
\ and\ \bibinfo {author} {\bibfnamefont {A.}~\bibnamefont {{McMurry}}}, 
\bibinfo {author} {\bibfnamefont {et}\ \bibnamefont {{al.}}},\ }
\href  {\doibase 10.1086/378512} {\bibfield  {journal} {\bibinfo  {journal} {\apj}\
  }\textbf {\bibinfo {volume} {599}},\ \unskip\ \bibinfo {pages}
  {626--660}, December (\bibinfo {year} {2003})}\BibitemShut {NoStop}%
\bibitem [{\citenamefont {{Zhao}}\ \emph {et~al.}(2015)\citenamefont {{Zhao}},
  \citenamefont {{Chen}}, \citenamefont {{Hartlep}},\ and\ \citenamefont
  {{Kosovichev}}}]{ZhaoJunwei.sunspot.waves.2015ApJ...809L..15Z}%
  \BibitemOpen
  \bibfield  {author} {\bibinfo {author} {\bibfnamefont {J.}~\bibnamefont
  {{Zhao}}}, \bibinfo {author} {\bibfnamefont {R.}~\bibnamefont {{Chen}}},
  \bibinfo {author} {\bibfnamefont {T.}~\bibnamefont {{Hartlep}}}, \ and\
  \bibinfo {author} {\bibfnamefont {A.~G.}\ \bibnamefont {{Kosovichev}}},\
  }\href {\doibase 10.1088/2041-8205/809/1/L15} {\bibfield  {journal} {\bibinfo
   {journal} {\apjl}\ }\textbf {\bibinfo {volume} {809}},\ p.\ \bibinfo {pages}
  {L15}, August (\bibinfo {year} {2015})}.
\end{thebibliography}
